\documentclass{PoS}

\usepackage{amsmath}
\usepackage{amssymb}
\usepackage{mathtools}
\usepackage{mathrsfs}
\usepackage{amsfonts}
\usepackage{bbm}
\usepackage{enumerate}
\usepackage{array}
\usepackage{booktabs}
\usepackage{textcomp}
\usepackage{caption}
\usepackage{cancel}
\usepackage{units}
\usepackage{multirow}
\usepackage{longtable}
\usepackage{relsize}
\usepackage{pdflscape}

\usepackage[latin1]{inputenc}
\usepackage{boxedminipage}
\usepackage{enumerate}
\usepackage[usenames,dvipsnames,table]{xcolor}
\usepackage{amssymb,latexsym}
\usepackage{graphicx}
\usepackage{subfigure}
\usepackage{multicol}
\usepackage{float}
\usepackage{slashbox}

\newcommand{\rep}[1]{\mathbf{#1}}
\newcommand{\crep}[1]{\overline{\mathbf{#1}}}

\newcommand{\zsix}[0]{\ensuremath{\mathbbm{Z}_{6-\text{II}}}}
\newcommand{\ztf}[0]{\ensuremath{\mathbbm{Z}_{2}\times\mathbbm{Z}_{4}}}
\newcommand{\ztt}[0]{\ensuremath{\mathbbm{Z}_{2}\times\mathbbm{Z}_{2}}}

\newcommand{\SO}[1]{\ensuremath{SO(#1)}}

\title{Lessons from an Extended Heterotic Mini-Landscape}

\ShortTitle{Lessons from an Extended Heterotic Mini-Landscape}

\author{Dami\'an Kaloni Mayorga Pe\~na\footnote{Speaker} , Paul-Konstantin Oehlmann\\
         Bethe Center for Theoretical Physics, \\
Physikalisches Institut der Universit\"at Bonn, Nussallee 12, 53115 Bonn, Germany\\
        E-mail: \email{damian}, \email{oehlmann@th.physik.uni-bonn.de}}

\abstract{We consider the key ingredients which made the \zsix\ Mini-Landscape a successful ground for model building. There one observes a preferred distribution for the fields of the standard model which has direct implications on the phenomenology, since it favors a heavy top quark and a plausible solution to the $\mu$-problem. We extend the exploration to the \ztf\ orbifold, where a similar pattern is identified. To illustrate our findings we present an explicit \ztf\ model that exhibits the MSSM spectrum plus vector-like exotics at the orbifold point. We take this construction as an example of a larger class of orbifold models which were anticipated at the GUT level. Furthermore we comment on VEV configurations to decouple the exotics while maintaining a meaningful $R$-symmetry which could forbid the $\mu$-term.
}
\FullConference{Proceedings of the Corfu Summer Institute 2012 "School and Workshops on Elementary Particle Physics and Gravity"\\
		September 8-27, 2012\\
		Corfu, Greece}

\begin{document}

\section{Introduction}
To date, string theory seems the most promising framework for an ultraviolet completion of the standard model of particle physics (SM) or any of its supersymmetric extensions. It is thus expected that all free parameters of the SM -such as masses and coupling constants- can be related to the compactification process from ten to four dimensions. Any reliable stringy completion must then reduce, in its low energy limit, to a field theory whose gauge group contains that of the SM ($G_{SM}$), in addition to three families of quarks and leptons, Higgses and the correct interaction terms.

We concentrate on orbifold compactifications \cite{Dixon:1985jw,Dixon:1986jc,Ibanez:1986tp} of the $E_8\times E_8$ heterotic string \cite{Gross:1984dd, Gross:1985fr}, commonly referred to as the \emph{heterotic brane world} \cite{Forste:2004ie,Nilles:2008gq}. A small part of that landscape has been analyzed, particularly in the case of the \zsix\ orbifold \cite{Kobayashi:2004ud,Kobayashi:2004ya} where a substantial class of models was found to be very successful for model building \cite{Buchmuller:2005jr,Buchmuller:2006ik, Lebedev:2006kn, Lebedev:2007hv, Lebedev:2008un}. The success of this so-called \emph{Mini-Landscape} is basically due to the following facts
\begin{enumerate}[(a)]
 \item The pattern of gauge symmetries makes the standard model gauge group to descend from an underlying $SO(10)$ or $E_6$, which remains unbroken at certain regions of the orbifold. The matter fields living at these locations will arise as complete GUT representations. This unification scheme is known as Local Grand Unification \cite{Ratz:2007my}, and provides an elegant solution for the doublet-triplet splitting problem commonly observed in standard GUTs.
\item There is a preferred distribution for the SM fields in the extra dimensions. Such distribution ensures a large top-Yukawa coupling \cite{Buchmuller:2006ik,Lebedev:2007hv}, a plausible solution to the $\mu$-problem \cite{Lebedev:2007hv,Casas:1992mk, Antoniadis:1994hg}, among other crucial requirements from the experimental point of view.
\end{enumerate}
Based on the results from the Mini-Landscape, we could establish a Zip-code for the location of SM fields in the extra dimensions. Our analysis has shown that
\begin{enumerate}[(i)]
\item the Higgs bosons $H_u$ and $H_d$ are free to propagate in the bulk, \label{uno}
\item the top quark lives in the bulk as well, whereas other members of the top family emerge from various locations in internal space \label{dos}
\item the first two families are located at points in extra dimensions and exhibit family symmetries \cite{Ko:2007dz,Kobayashi:2006wq}. \label{tres}
\end{enumerate}
We have also constructed heterotic models based on the \ztf\ orbifold: Having the complete classification of models allowed by this geometry \cite{Pena:2012ki}, we focused on those containing a GUT factor (such as $SO(10)$ or $E_6$). The breakdown of the GUT factor can be achieved by turning on discrete Wilson line (WL) backgrounds. Nevertheless the GUT level is enough to get an idea on the field distribution. The exploration led us to a refined selection of models where it seemed likely for conditions (\ref{uno}) to (\ref{tres}) to be satisfied.

Still, the full spectrum can only be determined after specifying the WLs, so that the explicit model construction is unavoidable. In this spirit, and aided by the C++ Orbifolder \cite{Nilles:2011aj}, we refined our selection of promising models by constructing the explicit Wilson lines consistent with the features which make the models to be so appealing. 

This work is intended to present the outcome of our search and it is organized in the following manner: Section \ref{zipcode} is devoted to a description of the Zip-code identified in the Mini-Landscape and some of its phenomenological implications. In section \ref{ztf} we shall review the construction and features of the \ztf, orbifold, the gauge groups it allows for and the relevant $SO(10)$ and $E_6$ models we found. We also reproduce the main lines of argumentation which lead us to conclude that an identical distribution of the SM fields is to be expected in this context. In section \ref{models} we describe which of the GUT models allow for a suitable Wilson line configuration and present an explicit realization. In section \ref{con} we shall conclude and summarize our findings.

\section{The Particle Zip-code in Mini--Landscape Models}
\label{zipcode}

Within the framework of the \zsix\ orbifold, hundreds of models have been identified to allow for a realistic MSSM structure. We concentrate our discussion on the models
of \cite{Lebedev:2007hv} as representatives for the Mini-Landscape, but before we discuss our findings let us briefly explain which kind of fields are allowed by the orbifold compactification. Depending on their localization properties we can distinguish between three classes of states:
\begin{itemize}

\item fields which are free to propagate in the 10-dimensional bulk

\item fields sitting at fixed points in the extra dimensions (representing ``3-branes'')

\item fields which can only propagate along fixed tori in compact space (representing ``5-branes'')

\end{itemize}

This distinction is relevant for our discussion, due to the observation that the geography of the fields in extra dimensions plays a crucial role in the low energy effective theory: In $4D$ we are aiming to achieve the gauge group of the SM (times some extra factors). From the orbifold point of view, such a gauge group originates as the intersection of all gauge factors which are characteristic to the various orbifold locations, in particular, there are some fixed points/tori where the gauge symmetry is enhanced to e.g. $SO(10)$ or even $E_6$. Since localized fields transform as complete representations of the local gauge group, it then follows that orbifold compactifications allow for the coexistence of complete and split multiplets of an underlying GUT. A similar situation is observed for the supersymmetries: fields in the bulk feel remnants of $\mathcal{N}=4$ SUSY, in contrast to those sitting at fixed tori and fixed points which experience remnants of $\mathcal{N}=2$ and $\mathcal{N}=1$, respectively. Arguments of this type motivate the idea that a particular distribution of the SM fields in the extra dimensions might explain the observed phenomenology. Bearing this in mind, we dedicate the remainder of this chapter to analyze the preferred locations for the SM fields observed in the Mini-Landscape as well as the physical implications of this peculiar distribution.

\subsection{The Higgs system}
\label{sec:higgs}

The minimal supersymmetric version of the SM contains one pair of Higgs superfields $H_u$ and $H_d$.
With regards to this vector-like pair, two important questions which raise from the phenomenological side need to be addressed: The first one is the doublet triplet splitting problem. As stressed before, local GUTs allow for certain fields to come in incomplete representations of the gauge symmetry, provided these fields originate from a certain region which does not experience the gauge enhancement. We expect the Higgses to live in those regions, so that the dangerous triplets which could accompany them, are definitely projected out of the spectrum.

The second situation which needs to be addressed here is the so called $\mu$-problem. Generally in string constructions one gets more than one pair of Higgses, so that one has to face the question of why all except one of them become heavy. For a large class of Mini-Landscape models we find that this problem is solved in a miraculous way \cite{Lebedev:2007hv,Kappl:2008ie}: In these models one pair of Higgs doublets lives at the untwisted sector. Such Higgses happen to be vector-like under all possible symmetries, and this is of great interest since any $R$-symmetry in the model will thus forbid a coupling of the form $\mu H_u H_d$ to be present in the superpotential $(\mathcal{W})$ \cite{Lebedev:2007hv,Kappl:2008ie,Lee:2011dya}. The $\mu$ term can be generated after SUSY breakdown. In particular, gravity mediation mechanisms relate the $\mu$ term directly to the gravitino mass $m_{3/2}$, as observed earlier in the context of field theoretic models \cite{Casas:1992mk}.

$R$-symmetries are crucial for this mechanism to work, and their origin is well understood: since the elements of the Lorentz group treat bosons and fermions in a different manner, any element of the extra dimensional Lorentz group $SO(6) \subset SO(9,1)$, which survives the compactification will introduce an $R$-symmetry in the $4D$ theory \cite{Nilles:2012cy,Bizet:2013gf}. 

The first lesson from the Mini-Landscape is then, that \emph{the Higgs pair
$H_u$ and $H_d$ should live in the bulk (untwisted sector)}. They correspond to gauge
fields in extra dimensions, compatible with the so-called gauge-Higgs
unification and represent continuous Wilson lines
as discussed in ref. \cite{Ibanez:1987xa,Forste:2005rs,Hebecker:2012qp}.

\subsection{The top quark}
\label{sec:top}

The mass of the top quark is of the order of the electroweak scale, one thus expects the gauge and the top-Yukawa couplings to be of the same order,
exhibiting gauge-top unification \cite{Hosteins:2009xk}. From the stringy point of view, these
couplings are given directly by the string coupling and we expect
the top quark Yukawa coupling at the trilinear level in the
superpotential. Given the fact that $H_u$ is a field in the
untwisted sector there remain only few allowed alternatives for the location of the top quark.

In the Mini-Landscape we find that, in most of the models, the left- and
right-handed top-multiplets belong to the untwisted sector
(bulk) and this is a guarantee for a sufficiently large Yukawa coupling. The
location of the other members of the third family is rather model
dependent, they are distributed over various sectors. Very often
the top quark Yukawa coupling is the only trilinear Yukawa
coupling one finds in the model.

This is the second lesson from the Mini--Landscape: \emph{left- and
right-handed top quark multiplets should be bulk fields}.

\subsection{The first two families of quarks and leptons}
\label{sec:families}
The Mini-Landscape models provide a grand unified picture
with families in the 16-dimensional spinor
representation of $SO(10)$ (in the case of $E_6$ enhancements, the families are expected to descend from the $\mathbf{27}$-plet). The analysis shows that three family models can only be achieved with at most two families as complete $SO(10)$ representations \cite{Buchmuller:2005jr}. Since the top-family will originate as a patchwork, the localized $\mathbf{16}$-plets correspond to the two light families, these live at fixed points of the orbifold twist. Due to their location, no trilinear coupling is allowed for these fields and for this reason quark-and lepton masses are suppressed. It is also observed that these families transform as a doublet under a $D_4$ family symmetry \cite{Ko:2007dz,Kobayashi:2006wq} inherited from the geometry. This symmetry can be used to avoid the problem of flavor
changing neutral currents. Thus we have a third lesson: \emph{the first two families are located at fixed points where the gauge symmetry is enhanced and enjoy of a non trivial transformation under a family symmetry}.

\subsection{The pattern of supersymmetry breakdown}
\label{susy}
The field configuration discussed in sects. \ref{sec:higgs} to \ref{sec:families} has many implications for the low energy, some of which have been already stressed. So far we have not discussed the breakdown of supersymmetry, which is also sensitive to the SM Zip-code. This discussion is more involved since it has to address the
question of moduli stabilization. Supersymmetry can be broken by gaugino condensation in the hidden sector \cite{Nilles:1982ik,Ferrara:1982qs}. The hidden groups one obtains in Mini-Landscape models are consistent with a gravitino mass in the (multi) TeV-range \cite{Lebedev:2006tr}
(provided the dilaton is fixed at a realistic grand unified gauge
coupling). If moduli stabilization proceeds along the lines of \cite{Kappl:2010yu,Anderson:2011cza}, we would then keep a run-away dilaton
and a positive vacuum energy. It can be shown that by adjusting the vacuum energy with a matter superpotential (downlifting
the vacuum energy)
one can fix the dilaton as well. The resulting picture \cite{Lowen:2008fm} is
reminiscent of a scheme known as \emph{mirage mediation} \cite{Choi:2004sx,Choi:2005ge,LoaizaBrito:2005fa}, at least
for gaugino masses and A-parameters: As bulk fields, gauginos experience remnants of $\mathcal{N}=4$ SUSY and due to that, their masses are suppressed by a factor of $\log ({M_{\rm Planck}/{m_{3/2}}})$ \cite{Choi:2007ka}. Scalar masses are more model
dependent and could be as large as the gravitino mass $m_{3/2}$ \cite{Lebedev:2006qq}. However, since Higgses and top quark are also bulk fields, the same suppression factor is expected for the mass of Higgsinos and the stop. Thus, \emph{one expects $m_{3/2}$
and scalar masses of the first two families
in the multi-TeV range while stops and Higgsinos are in the TeV range.}
This is a result of the location of fields in the
extra dimensions and provides the fourth lesson of the Mini--Landscape \cite{Krippendorf:2012ir}.

\section{The $\boldsymbol{\mathbbm{Z}_{2}\times\mathbbm{Z}_{4}}$ Orbifold}
\label{ztf}
Distinctive of the heterotic string is the obvious dimensional mismatch between left- and right-movers: Whereas the left-moving modes are 26 bosonic degrees of freedom, the right moving modes are taken to be those of the $10D$ superstring. Such a mismatch is cured by compactification of the $16$ additional left-movers, and depending on the compactification, one ends up with either $E_8\times E_8$ or $SO(32)$ as the gauge group for the theory.

Out of the ten remaining bosonic coordinates, six need to be compactified in order to retain an effective four dimensional theory. The compactification of our interest proceeds in the following manner: For convenience we complexify the extra dimensional coordinates, so that one deals with $\mathbbm{C}^3$ instead of $\mathbbm{R}^6$. One takes the lattice $\Gamma$ spanned by the simple roots of $SU(2)^2\times SO(4)\times SO(4)$ in which each of the subfactors aligns on a different plane. The orbifold is obtained by modding out a \ztf\ lattice automorphism (point group) out of the torus $\mathbbm{C}^6=\mathbbm{C}^3/\Gamma$.

The generators of the \ztf\ point group act on $\mathbbm{C}^3$ according to the prescription
\begin{equation}
 \vartheta_{k}=(e^{2\pi {\rm i}v_k^1}, e^{2\pi {\rm i}v_k^2}, e^{2\pi {\rm i}v_k^3}), \,\,\, k=2,4\, ,
\end{equation}
where $v_k\equiv(0,v_k^1,v_k^2,v_k^3)$ are known as twist vectors. The $\mathbbm{Z}_2$ generator $\vartheta_2$ is taken as a simultaneous rotation of $\pi$ in the first and third complex plane, while $\vartheta_4$ corresponds to rotations of $\pi/2$ and $3\pi/2$ in the last two planes. The twists are thus given by
\begin{align*}
\textstyle v_2=(0,\frac{1}{2},0,-\frac{1}{2}) \quad \text{and} \quad v_4=(0,0,\frac{1}{4},-\frac{1}{4}) \, .
\end{align*}
Let us now discuss the features of the string theory compactified on such background. The choice of the twist ensures that we end up with one surviving supersymmetry in $4D$. When it comes to the massless spectrum, one can distinguish between twisted and untwisted states depending on whether or not a twist is needed for the strings to close. Examples of untwisted states are the SUGRA multiplet, the vector multiplets describing the gauge interactions and some additional chiral fields which are free to propagate in the bulk. Twisted states close according to the following boundary condition in the internal space ($Z\in\mathbbm{C}^3$)
\begin{equation}
Z(\tau,\sigma+\pi)=\vartheta_{2}^n \vartheta_{4}^m Z(\tau,\sigma)+n_\alpha \text{e}_\alpha ,\label{bc}
\end{equation}
with $\{\text{e}_\alpha\}_{\alpha=1}^6$ being a basis for $\Gamma$. Since the twist $\vartheta_{2}^n \vartheta_{4}^m$ closes the string, the resulting states is said to belong to the twisted sector $T_{(m,n)}$. After computing the mode expansion for $Z$ one can observe that the string center of mass must lie at any of the fixed points of $\vartheta_{2}^n \vartheta_{4}^m$. A detailed description of the twisted sectors for this orbifold can be found in ref. \cite{Pena:2012ki}.

For the gauge coordinates one imposes the following condition
\begin{equation}
X^I(\tau,\sigma+\pi)=X^I(\tau,\sigma)+2\pi (nV_2+mV_4+n_\alpha W_\alpha)^I  \,,\quad I=1,...,16\, ,
\end{equation}
since the embedding of the twist into the gauge coordinates is necessary to ensure the modular invariance (MI) of the vacuum to vacuum amplitude. The embedding we have constructed by introducing the shifts $V_2$ and $V_4$ and the discrete Wilson lines $W_\alpha$, which are sixteen dimensional vectors and mimic the roto-translation from eq. \eqref{bc}. We avoid most of the technicalities about the massless spectrum\footnote{For details on these matters we refer the reader to \cite{Bailin:1999nk,Choibook,Vaudrevange:2008sm,RamosSanchez:2008tn}}. Nonetheless, it is important to remark that the spectrum of the theory is entirely determined by the choice of the embedding vectors. The same is true for the gauge group and the corresponding interactions. Our model building strategy intends to identify a certain set of models in which a GUT structure such as $E_6$, $SO(10)$ or $SU(5)$ is possible, this is expected to occur at the level in which all discrete Wilson lines are switched off. The WLs are then used to break the GUT factor down to the SM gauge group. The right matter content is expected to arise at that level, accompanied of certain vector-like exotics, likely to decouple from the low energy regime.
\subsection{Embedding Classification}
As a first step towards model building in this orbifold, we are interested in the gauge groups that the geometry allows for. It was already pointed out that the gauge group of the model is set by the choice of the gauge embedding, hence our intention is to classify all embeddings allowed by this geometry. The choice of the vectors $V_2$, $V_4$ and $W_\alpha$ is not entirely arbitrary: On one hand, there are some requirements arising from modular invariance (see e.g. ref. \cite{Ploger:2007iq}). On the other hand, consistency of the embedding with the orbifold geometry constraints the order of the vectors and relates between the Wilson lines.
\begin{enumerate}[(i)]
 \item Since the generators $\vartheta_2$ and $\vartheta_4$ are of order 2 and 4 respectively, i.e. the vectors $2V_2$ and $4V_4$ must belong to the $E_8\times E_8$ lattice.
 \item Similarly, the Wilson lines have to be all of order 2. In addition to that, the point group elements relate between some of the Wilson lines: Take $\text{e}_{2i-1}$ and $\text{e}_{2i}$ to span the $i$-th complex plane. Since the $\mathbbm{Z}_4$ generator relates $\text{e}_{3}$ and $\text{e}_{4}$, the Wilson lines $W_3$ and $W_4$ can differ at most by an $E_8\times E_8$ lattice vector. The same argument applies for $W_5$ and $W_6$.
\end{enumerate}
In contrast to the Wilson line backgrounds, the shift vectors $V_2$ and $V_4$ are necessarily non trivial. Since we are interested in obtaining all inequivalent models, the simplest alternative is to assume all WLs are off and search for consistent pairs $(V_2,V_4)$. Crucial for our exploration is that if two embeddings differ by lattice automorphisms or certain lattice vector combinations, the models they lead to are identical. This implies that we can zoom in a particular region of the lattice, and from this region we can find all possible models. We take the compactification lattice for the gauge space as that of $E_8\times E_8$. This choice is motivated by the flexibility of $E_8$ when it comes to nest a grand unified group. The direct product structure eases the search for inequivalent vectors, because we can take the shifts as the direct product of two eight dimensional vectors and study them independently. For $V_4$  we took the ten inequivalent shifts found for the $\mathbbm{Z}_4$ orbifold \cite{Katsuki1989}, with these quantities at hand we explored the corresponding $\mathbbm{Z}_2$ shifts consistent with each of the $V_4$'s. 

In order to avoid constructing the $E_8$ automorphism group we made use of the Weyl reflections to relate between lattice vectors. We took the set of reflections leaving the vectors invariant and used this as a basis to generate the stabilizer of $V_4$ within $\text{Aut}(\mathfrak{e}_8)$; finally we employed the stabilizer elements to relate between the $V_2$ shifts\footnote{There is a little drawback in this algorithm since the Weyl reflections from the stabilizer are in general not enough to generate any arbitrary automorphism which leaves $V_4$ intact. Fortunately, the output is reduced enough so that when one has emebeddings which are potentially equivalent, one can check by hand if there is an automorphism relating between them.}.

It can be shown that the gauge group of a certain embedding does not change if one adds lattice vectors to the shifts. As for now we are only interested in the diversity of gauge groups \ztf\ permits, we take any given pair of vectors as equivalent if they differ by a lattice translation. As the $V_2$ shifts must be half of a lattice vector, it suffices to consider all half lattice vectors of $E_8$ whose norm is smaller than two. After this we are left with $75$ combinations for a single $E_8$. These combinations are then paired together to form sixteen dimensional vectors. Only those pairings which agree with modular invariance are useful for model building. Allowed embeddings must satisfy
\begin{align}
2\left(V_2^2-\frac{1}{2}\right)&=0\mod 2 \, , \label{1mod}\\
4\left(V_4^2-\frac{1}{8}\right)&=0\mod 2 \, , \label{2mod}
\end{align}
in addition to a mixed constraint\footnote{Our construction misses the effects of the lattice vectors, hence the MI condition which mixes between $V_2$ and $V_4$ \cite{Ploger:2007iq} gets milder. Embeddings which can be made modular invariant necessarily have to satisfy eq. \ref{weak}}
\begin{equation}
4\left(V_2\cdot V_4+\frac{1}{8}\right)=0\mod 1\,. \label{weak}
\end{equation}
We found 144 combinations which survive these conditions and the gauge groups for these models were computed. Among the relevant features of the gauge structures we found, it is worth to highlight the large number of embeddings in contrast to \zsix\ where only 61 are possible. In our model, 35 embeddings contain an $SO(10)$ factor, we also have 26 $E_6$ models and 25 containing a $SU(5)$ factor. For the \zsix\ one has 14 $SO(10)$ models, 16 with $E_6$ and 4 with $SU(5)$.

\subsection{Gauge Topography}

To judge the fertility of the shift embeddings discussed in the previous section, we can compute the spectrum at the GUT level and then analyze the possible effects of the Wilson lines. With this information one can implement a strategy to retain three families in the low energy.

At the GUT level, the multiplicities for the twisted states from $T_{(0,1)}$, $T_{(0,3)}$ and $T_{(1,3)}$ are equal to the number of fixed points/tori in each of the sectors. However, the sectors $T_{(1,0)}$, $T_{(0,2)}$ and $T_{(1,2)}$ contain some \emph{special} fixed points which get identified to each other by the $\mathbbm{Z}_4$ generator. At these points there is an enhancement of the gauge symmetry, as a consequence, fields at those locations will furnish complete representations of the enhanced gauge group. In those sectors, the fields sitting at normal fixed points will also be present at the special ones, but at those special locations some additional states appear.

When it comes to the effects of the Wilson lines, there are some differences in the way they act on matter fields. In the untwisted sector, for instance, the states are sensitive to the Wilson lines, as expected for fields which propagate in the bulk, no new states are introduced and out of those found at the GUT level only those which survive the WL projections will appear in the spectrum.

The Wilson lines act differently on the fixed points and hence split the multiplicities for the states in a given twisted sector. Thus, the Wilson lines are not only needed for the breaking of the GUT factor, but must also allow for a distribution in which the SM matter comes in three copies. Depending on the manner the WL configuration acts on the states sitting there, the singularities fit in the following classification:
\begin{enumerate}[(i)]
 \item At some fixed points/tori, the Wilson lines act trivially and hence, the matter sitting at those fixed points is the same as that one had at the GUT level. These singularities are called \emph{protected} and they are the only locations in the orbifold where the twisted states furnish complete representations under the GUT group.
 \item At the \emph{split} singularities, the states are the same as those at the GUT level, but in this case the Wilson lines act as projectors, so that only some pieces of the GUT multiplets will survive. This is very similar to what occurs in the untwisted sector.
\item When the Wilson lines enter the mass equations for the states sitting at a certain singularity, the spectrum at this location is entirely disconnected from that at the GUT level. Such singularities we call \emph{unshielded} because the matter sitting there is extremely sensitive to the specific choice of the WLs.
\end{enumerate}
\
Clearly, we can use the spectrum at the GUT level to infer what kind of matter is likely to appear at a certain point. Unfortunately this is only true for the untwisted sector, protected and split singularities. We must remark that the matter states living at unshielded singularities can only be found after specifying the Wilson lines. To avoid an exhausting search for suitable Wilson lines we assume that the relevant fields of the MSSM do not sit at unshielded locations. If we want for the $SO(10)$ models the families to be complete GUT representations, they must then arise from protected fixed points. If we consider $E_6$ models both split and protected singularities are favored to allocate a whole family, provided the existence of a Wilson line configuration which locally breaks the gauge group to $SO(10)$, while leaving the $\mathbf{16}\subset\mathbf{27}$ untouched. In all other cases the families will arise as a patchwork of states originating from various locations in the orbifold.

It was previously pointed out that for any of the 144 shift embeddings allowed for \ztf, one can add certain lattice vectors without spoiling modular invariance. In particular, for each model we will obtain two inequivalent \emph{brother} spectra. The addition of lattice vectors does not modify the gauge group nor the untwisted matter content so that brother models only differ by the twisted matter. We computed the spectra for all of those shifts leading to an $SO(10)$ or an $E_6$ factor. This information can help us to find out which are the fertile places for the fields to sit, i.e. the locations where $\mathbf{16}$- or $\mathbf{27}$-plets appear in a given model. Then we use this information to decide what are the directions along which we can switch on the Wilson lines in order to get at most\footnote{In those cases where less than three complete GUT multiplets are present, the members of the remaining families are expected to arise from unshielded or split singularities.} three (net)surviving $\mathbf{16}$-plets ($\mathbf{27}$'s).

\section{Promising Candidates}
\label{models}
In addition to the three families of the SM, we have to ensure that the interactions of the standard model are reproduced in an accurate manner. In general terms, one of the $U(1)$'s from the model is anomalous. Such anomaly is canceled via the $4D$ Green Schwarz mechanism, and the large FI term associated to it is counter accounted by assigning VEVs to certain SM singlets. As already pointed out, the large mass for the top quark in comparison to other SM fields, hints to its special stringy origin: Whereas all other Yukawas arise from $L$-point couplings with $L-3$ VEV fields, the one for the top arises at trilinear order. This implies that the presence of a heavy top in a given model can be checked before dealing with VEV configurations, particularly, if we assume none of the pieces involved in the top coupling comes from an unshielded singularity, the trilinear couplings must also be present at the GUT level. An operator of the form $\mathbf{16}\cdot\mathbf{16}\cdot\mathbf{10}$ (or $(\mathbf{27})^3$) allowed in an $SO(10)$ (or $E_6$) model, will induce the desired $\overline{U}QH_u$, if the relevant pieces survive the Wilson line projections.

The allowed couplings in the superpotential can be inferred from the orbifold CFT \cite{Dixon:1986qv,Hamidi:1986vh, Stieberger:1992bj,Font:1988mm, Kobayashi:2011cw}. From there we know that, in addition to the requirement for gauge invariance, some properties of the internal space will manifest as symmetries in the field theory. On one hand, those elements from the internal Lorentz group which are consistent with orbifolding will induce discrete $R$-symmetries. On the other hand, the orbifold geometry will introduce some discrete non-$R$ symmetries (which in the absence of Wilson lines may enhance to non-Abelian factors). The point group is one example of such symmetries: A state from the $T_{(m,n)}$ sector transforms with charges $m$ and $n$ under the $\mathbbm{Z}_2$ and $\mathbbm{Z}_4$ generators, respectively. One can thus see that, among all possible sectors for the fields to originate from, the point group symmetry will leave us only with the following alternatives for a trilinear coupling\footnote{In general, the point group itself gives more possibilities for a trilinear coupling, however
all those involving $T_{(1,1)}$ are not viable since this sector does not contain any left-chiral state. One the other hand, $R$ charge conservation forbids couplings such as $T_{(0,1)}T_{(0,1)}T_{(0,3)}$ which share a common fixed torus \cite{Casas:1991ac}.}
\begin{equation*}
\begin{array}{lclcl}
1. \, \,\, \, UUU\,, & \,\,\,\,\,\,\,\,\, \,\,\,\,\,\,\,\,\, & 2.  \, \,\, \, T_{(0,2)} T_{(0,2)} U\,, &\,\,\,\,\,\,\,\,\, \,\,\,\,\,\,\,\,\,& 3.  \, \,\, \,T_{(1,0)} T_{(1,0)} U\,,\\
4. \, \,\, \, T_{(1,2)} T_{(1,2)} U\,, & & 5.  \, \,\, \, T_{(0,1)} T_{(0,3)} U\,, & & 6. \, \,\, \, T_{(0,2)} T_{(1,2)}T_{(1,0)}\,, \\
7.\, \,\, \,T_{(1,3)} T_{(1,3)}T_{(0,2)}\,, &  & 8. \, \,\, \, T_{(1,3)} T_{(0,3)}T_{(1,2)}\,, & & 9.  \, \,\, \,T_{(1,3)} T_{(0,1)}T_{(1,0)}\,.
\end{array}
\end{equation*}
Now we use the spectra to determine which of the previous couplings is supported by any of the models. In ref. \cite{Pena:2012ki}, the effects of the various Wilson lines were analyzed. We consider the spectrum in combination with the schematic action of the Wilson lines, with this information we intend to determine which models support any of the couplings depicted above. For conciseness, and since both $SO(10)$ and $E_6$ models feature similar properties, we present our findings for $SO(10)$ and defer the discussion about $E_6$ to the appendix \ref{e6}. Among all $SO(10)$ embeddings only one was found to allow for three complete $\mathbf{16}$-plets. In that model, all of the $\mathbf{16}$-pets belong to the $T_{(1,2)}$ sector, so that the top-Yukawa must arise from a coupling of the form $T_{(1,2)} T_{(1,2)} U$. The untwisted sector contains two $\mathbf{10}$-plets; unfortunately, none of them permits a gauge invariant coupling of the form $\mathbf{16}\cdot\mathbf{16}\cdot\mathbf{10}$. Hence, \emph{for the \ztf\ orbifold we constructed there is no $SO(10)$ model with three complete GUT families and a large top-Yukawa coupling.}

Configurations with only two complete families are more commonly found. These families usually sit at the same twisted sector and transform among each other due to an underlying $D_4$ flavor symmetry, which, as pointed out previously, is a consequence of a Wilson line being off. By looking at the spectra for these models one can search for allowed operators of the form $\mathbf{16}\cdot\mathbf{16}\cdot\mathbf{10}$ which could involve twisted fields. Surprisingly, couplings of this kind are not found, and hence the only alternative which is left for a trilinear coupling at the GUT level is $UUU$. Three embeddings were found to contain an $SO(10)$ factor, with two complete families and a purely untwisted trilinear interaction; the remainder of this section is devoted to discuss them in detail.\\

The first promising embedding is realized by the vectors

\begin{align}
& {\textstyle V_2^{SO(10),1}=(\frac12, \frac12,1,0,0,0,0,0)(1,1,1,0,0,0,0,0)}\,, \nonumber \\
& {\textstyle V_4^{SO(10),1}=(\frac12, \frac14, \frac14, 0,0,0,0,0)(\frac12,\frac12,\frac12,0,0,0,0,0)}\,, \label{1}
\end{align}
which lead to the gauge group $[SO(10)\times U(1)^3]\times[SO(10)\times SU(4)]$ (the squared brackets are set to distinguish between the original $E_8$ factors). The first $SO(10)$ factor is the relevant one and the twisted $\mathbf{16}$-plets appear at $T_{(1,2)}$. In order to achieve only two protected fixed tori in this sector one must have $W_3\neq 0$, $W_5=0$ and either $W_1=0$ or $W_2=0$ but not both.\\

The second embedding corresponds to
\begin{align}
& \textstyle V_2^{SO(10),2} = (2,0,0,0,0,-1,0,0)(\frac32,0,-\frac12,-\frac12,-\frac12,-\frac12,0,\frac12)\,,\nonumber \\
& \textstyle V_4^{SO(10),2}= (1,0,0,0,0,0,0,0)(\frac34,\frac14,0,0,0,0,0,0)\,, \label{2}
\end{align}
its gauge group is $[SO(14)\times U(1)]\times [SO(10)\times U(1)^3]$, while the third one is generated by the shifts
\begin{align}
& \textstyle V_2^{SO(10),3}=(1,-\frac12,0,0,0,-\frac12,0,0)(\frac54,-\frac14,\frac34,\frac34,\frac34,\frac34,-\frac14,\frac14)\,,\nonumber \\
& \textstyle V_4^{SO(10),3} = (\frac12,0,0,0,0,0,0,0)(\frac54,-\frac14,-\frac14,-\frac14,-\frac14,-\frac14,\frac12,-\frac12)\,, \label{3}
\end{align}
leading to the gauge symmetry $[SO(10)\times SU(2)^2\times U(1)]\times [SU(8)\times U(1)]$. For the last configurations, three Wilson lines are needed in order to achieve two protected $\mathbf{16}$-plets in the $T_{(1,3)}$ sector. The Wilson line which remains off can be either $W_1$ or $W_2$.\\
\\
As already pointed out, the presence of a trilinear Yukawa is somehow a necessary requirement. In models with two complete $SO(10)$ families the desired Yukawa can be achieved if the up-type Higgs and left- and right- handed top quark are untwisted fields. However, there is a little drawback to overcome: we need to find a Wilson line configuration which ensures the coupling $\overline{U}QH_u$ $(\subset \mathbf{16}\cdot\mathbf{16}\cdot\mathbf{10})$ survives the projections. We have developed a search strategy which favors certain embeddings depending on their potential features, but we must not forget that these features can be spoiled by the Wilson lines. We have then arrived at a stage where we need to actually construct WL backgrounds to prove that our strategy works.

\subsection{An Explicit $SO(10)$ Example}
\label{explicit}
\begin{figure}[htb]
\centering
\includegraphics[width=11.25cm]{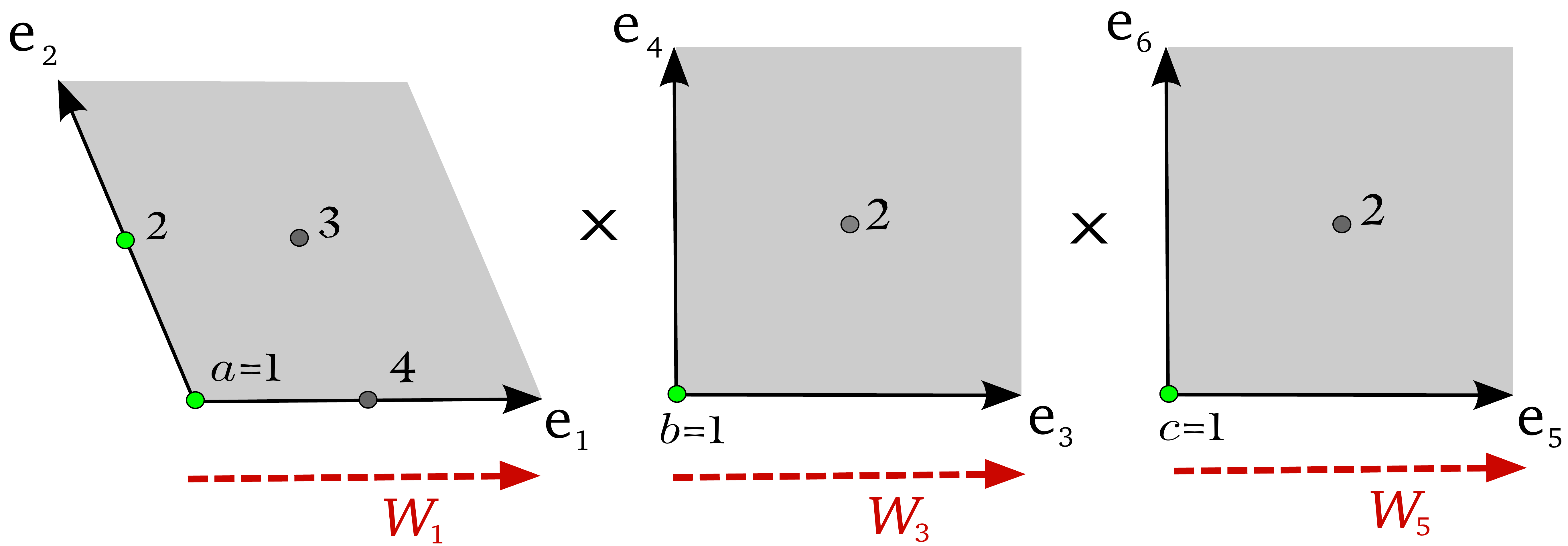}
\caption{The fixed points of the $T_{(1,3)}$ sector. Note that for the model under consideration and in the case $W_{1,3,5}\neq 0$, the $\mathbf{16}$-plets living at the fixed points $a=1,2$ ($b=1,c=1$) are not affected by the WL projections.}
\label{families}
\end{figure}

The embeddings previously discussed fit the Mini-Landscape Zip-code, at least at the GUT level. If such a picture can be retained after switching on the Wilson lines, one would expect the phenomenology of these models to develop along the lines of section \ref{zipcode}. Here we explore one model and present a WL configuration which partially agrees with our expectations. We took the model defined by the vectors in eq. \eqref{3}, whose complete spectrum at the GUT level is presented in table \ref{tab:BenchmarkSpectrum}.
\begin{table}[h!]
\renewcommand{\arraystretch}{1.15}
\begin{center}
\begin{tabular}{l c r}
\begin{tabular}{|l|l|} \hline
\multirow {7}*{$U$} & 1  $( \mathbf{16},  \mathbf{2},  \mathbf{1},  \mathbf{1})_{0,   -1}$\\
    &  1  $( \mathbf{16},  \mathbf{1},  \mathbf{2},  \mathbf{1})_{ 0,    1} $\\
  & 1 $( \mathbf{10},  \mathbf{2},  \mathbf{2},  \mathbf{1})_{ 0,    0}$\\
  &  1 $(  \mathbf{1},  \mathbf{1},  \mathbf{1},  \mathbf{1})_{-12,    0}$\\
  &  1 $(  \mathbf{1},  \mathbf{1},  \mathbf{1},  \mathbf{1})_{ 12,    0}$\\
  &  1 $(  \mathbf{1},  \mathbf{1},  \mathbf{1}, \mathbf{28})_{ 6,    0}$\\
  & 1 $(  \mathbf{1},  \mathbf{1},  \mathbf{1},\overline{\mathbf{28}})_{6,    0}$\\  \hline
\end{tabular}
&
\begin{tabular}{|l|l|}\hline
\multirow{2}*{$T_{(0,1)}$}   &  4 $(  \mathbf{1},  \mathbf{1}, \mathbf{1},  \mathbf{8})_{6,    1}$ \\
       & 4 $ (  \mathbf{1},  \mathbf{1},  \mathbf{1}, \overline{\mathbf{8}})_{    0,    1}$  \\ \hline
\multirow{4}*{$T_{(0,2)}$}         & 10 $(  \mathbf{1},  \mathbf{2},  \mathbf{2},  \mathbf{1})_{6,    0} $\\
   &10 $( \mathbf{10},  \mathbf{1},  \mathbf{1},  \mathbf{1})_{  -6,    0}$ \\
   & 6 $(  \mathbf{1},  \mathbf{1}, \mathbf{1},  \mathbf{1})_{   -6,   -2}$ \\
    &6 $(  \mathbf{1},  \mathbf{1},  \mathbf{1},  \mathbf{1})_{   -6,    2}$  \\ \hline
\end{tabular}
&
\begin{tabular}{|l|l|}\hline
  \multirow{2}*{$T_{(0,3)}$} & 4 $(  \mathbf{1},  \mathbf{1},  \mathbf{1}, \overline{\mathbf{8}})_{ 6,   -1}$ \\
    & 4 $(  \mathbf{1},  \mathbf{1},  \mathbf{1},  \mathbf{8})_{ 0,   -1}  $ \\ \hline
$T_{(1,0)}$ & 4 $(  \mathbf{1},  \mathbf{1},  \mathbf{2},  \mathbf{8})_{-3,    0} $ \\ \hline
$T_{(1,1)}$ & \multicolumn{1}{c|}{-}  \\ \hline
    $T_{(1,2)}$  &    4 $(  \mathbf{1},  \mathbf{2},  \mathbf{1}, \overline{\mathbf{8}})_{  -3,    0} $  \\ \hline
  \multirow{3}*{$T_{(1,3)}$} & $16 ( \mathbf{16},  \mathbf{1},  \mathbf{1},  \mathbf{1})_{    3,    0}$ \\
 & 16 $(  \mathbf{1},  \mathbf{2},  \mathbf{1},  \mathbf{1})_{  3,    1 } $ \\
   & 16 $(  \mathbf{1},  \mathbf{1},  \mathbf{2},  \mathbf{1})_{ 3,   -1} $ \\ \hline
\end{tabular}
\end{tabular}
\caption{Matter spectrum corresponding to the shift embedding given in eq. (4.3). For each state, the bold numbers label its corresponding representation under $SO(10)\times SU(2)\times SU(2)\times SU(8)$ respectively, whereas the subindices label the charges under the $U(1)$ symmetries.
}
\label{tab:BenchmarkSpectrum}
\end{center}
\end{table}\\
At this stage one can see that the untwisted sector allows for the gauge and space group invariant coupling
\begin{align*}
( \mathbf{16},  \mathbf{2},  \mathbf{1},  \mathbf{1})_{0,   -1}\cdot ( \mathbf{16},  \mathbf{1},  \mathbf{2},  \mathbf{1})_{ 0,    1}\cdot ( \mathbf{10},  \mathbf{2},  \mathbf{2},  \mathbf{1})_{ 0,    0}\,,
\end{align*}
which in fact is allowed by all CFT selection rules. This is the kind of coupling from which we expect the top-Yukawa to originate from. We can also observe that in the $T_{(1,3)}$ sector we get 16 identical copies of the representation $( \mathbf{16},  \mathbf{1},  \mathbf{1},  \mathbf{1})_{    3,    0}$, one sitting at each of the fixed points. Out of these, two will remain intact after the Wilson line configuration is set on. Aided by the C++ Orbifolder we searched for MI alternatives for $W_1$, $W_3$ and $W_5$ (see fig. \ref{families}). The Wilson line configuration is expected to break the $SO(10)$ factor down to the desired $G_{SM}$, we are looking for embeddings which allow for three net generations plus vector-like exotics and the standard hypercharge generator $U(1)_Y\subset SU(5)$ to be non-anomalous. As an outcome of this exploration, roughly 200 inequivalent choices for the Wilson line configuration were found to comply with our requirements. A similar amount of models is found for the two other shift embeddings presented before. To illustrate our findings let us study the effects of the following choice for the Wilson lines
\begin{align*}
W_1 &=  \textstyle ( -\frac12,   \frac12,  -\frac32,  -\frac12,     0,    -1,    -1,     2)  (-\frac34,  -\frac74,  -\frac14,  -\frac14,  -\frac14,   \frac74,   \frac34,   \frac34) \, , \\
W_3 &=  \textstyle (   -1,   \frac32,  -\frac12,   \frac12,     0,   \frac12,     0,     2)  (
-\frac12,     1,    -2,     0,   \frac32,    -1,  -\frac32,  -\frac32) \, ,\\
W_4 &=  \textstyle ( -\frac54,   \frac54,   \frac14,  -\frac14,   \frac34,  -\frac14,   \frac54,   \frac94)  (
0,     1,     1,     2,    -1,  -\frac12,     2,   \frac32) \,.
\end{align*}
This breaks the gauge group to $G_{SM}\times G_{\rm Hidden}\times U(1)^9$, where $G_{\rm Hidden}=SU(3)\times SU(2)$ is a subgroup from the $SU(8)$ in the second $E_8$. As expected one of the $U(1)$'s is anomalous, but the hypercharge generator is orthogonal to it. The splitting for the relevant untwisted fields at the GUT level, in terms of $G_{SM}$ is given by
\begin{align}
( \mathbf{10},  \mathbf{2},  \mathbf{2},  \mathbf{1})_{ 0,    0} &\rightarrow \overbrace{(\mathbf{1},\mathbf{2})_{-1/2,\cdots}}^{H_u}+\overbrace{(\mathbf{1},\mathbf{2})_{1/2,\cdots}}^{H_d} \,, \nonumber\\ \nonumber\\
( \mathbf{16},  \mathbf{2},  \mathbf{1},  \mathbf{1})_{0,   -1} &\rightarrow (\mathbf{1},\mathbf{1})_{-1,\cdots}+\underbrace{(\overline{\mathbf{3}},\mathbf{1})_{1/3,\cdots}}_{U} \,, \label{split}\\
( \mathbf{16},  \mathbf{1},  \mathbf{2},  \mathbf{1})_{ 0,    1}&\rightarrow \underbrace{(\mathbf{3},\mathbf{2})_{-1/6,\cdots}}_{Q} \nonumber\, ,
\end{align}
where the dots stand for additional $U(1)$ charges different than the hypercharge, the previous splitting implies that the trilinear couplings $\overline{U}HQ$ is retained. The spectrum after the Wilson line breaking is presented in table \ref{tab:SpectrumSM}, in addition to the protected $\mathbf{16}$-plets from $T_{(1,3)}$, one observes the presence of additional triplets coming from other locations, note however that there are exactly three $(\mathbf{3},\mathbf{2})_{\frac16}$, $(\overline{\mathbf{3}},\mathbf{1})_{-\frac23}$ and $(\mathbf{1},{\mathbf{1}})_{-1}$. The spectrum contains four fields of the form $(\mathbf{1},\mathbf{2})_{-\frac12}$ (including the Higgs $H_u$ introduced in eq. \eqref{split}), hence we have only one Higgs pair for this model. When it comes to the down-type quarks we find 6 extra $(\overline{\mathbf{3}},\mathbf{1})_{-\frac13}$ together with their complex conjugates. Additional exotics are observed, they come in vector-like pairs with respect to the SM and we expect them to decouple after assigning VEVs to some of the SM singlets $n_i$, $s_i$, $\tilde{v}_i$ and $\overline{s}_i$. We must insist that the choice of Wilson lines presented here serves merely to illustrate that it is possible to find Wilson lines leading to a three family model. However, one can see that this model can not follow the pattern depicted in section \ref{susy}. To see this note that states like $\chi_i$ or $m_i$ lack of a conjugate counterpart, so that they can only be decoupled from the spectrum in the case the VEV configuration breaks all non-Abelian factors in the hidden group, leaving no chances for gaugino condensation to occur.

\begin{table}[h!]
\renewcommand{\arraystretch}{1.15}
\begin{center}
\begin{tabular}{|c|c|c||c|c|c|}\hline
\# & Rep. & label & \# & Rep. & label\\ \hline
3 & $(\overline{\mathbf{3}},\mathbf{1},\mathbf{1},\mathbf{1})_{\frac23}$ & $\overline{u}$ & 69 & $(\mathbf{1},{\mathbf{1}},\mathbf{1},\mathbf{1})_0$ & $n$ \\
3 & $(\mathbf{1},{\mathbf{1}},\mathbf{1},\mathbf{1})_{-1}$ & $\overline{e}$ & 32 & $(\mathbf{1},{\mathbf{1}},\mathbf{1},\mathbf{1})_{-\frac12}$ & $ r $ \\
3 & $(\mathbf{3},\mathbf{2},{\mathbf{1}},{\mathbf{1}})_{-\frac16}$ & $q$ & 4 & $(\mathbf{1},{\mathbf{1}},\mathbf{1},\mathbf{2})_{-\frac12}$  & $ b $ \\
4 & $(\mathbf{1},\mathbf{2},\mathbf{1},\mathbf{1})_{\frac12}$ & $l$ & 30 & $(\mathbf{1},{\mathbf{1}},\mathbf{1},\mathbf{1})_{\frac12}$ & $\overline{r}$ \\
1 & $(\mathbf{1},\mathbf{2},\mathbf{1},\mathbf{1})_{-\frac12}$ & $\overline{l}$ & 4 & $(\mathbf{1},{\mathbf{1}},\overline{\mathbf{\mathbf{3}}},\mathbf{1})_{0}$ & $s$ \\
9 & $(\overline{\mathbf{3}},\mathbf{1},\mathbf{1},\mathbf{1})_{-\frac13}$ & $\overline{d}$ & 10 &  $(\mathbf{1},{\mathbf{1}},\mathbf{1},\mathbf{2})_{0}$ & $\tilde{v}$\\
6 & $(\mathbf{3},\mathbf{1},\mathbf{1},\mathbf{1})_{\frac13}$ & $d$ & 8  & $(\mathbf{1},{\mathbf{1}},\mathbf{3},\mathbf{1})_{0}$ & $\overline{s}$ \\
6 & $(\mathbf{3},{\mathbf{1}},\mathbf{1},\mathbf{1})_{-\frac16}$ & $f$ & 2 & $(\mathbf{1},{\mathbf{1}},\overline{\mathbf{3}},\mathbf{1})_{\frac12}$ & $\chi$\\
8 & $(\mathbf{1},{\mathbf{2}},\mathbf{1},\mathbf{1})_{0}$ & $v$ & 5 & $(\mathbf{1},{\mathbf{1}},\mathbf{1},\mathbf{2})_{\frac12}$ & $\overline{b}$\\
1 & $(\mathbf{3},{\mathbf{1}},\mathbf{1},\mathbf{2})_{-\frac16}$ & $m$ & 2 & $(\mathbf{1},{\mathbf{1}},\overline{\mathbf{3}},\mathbf{1})_{-\frac12}$ & $\tilde{\chi}$\\
8 & $(\overline{\mathbf{3}},\mathbf{1},\mathbf{1},\mathbf{1})_{\frac16}$ & $\overline{f}$ & & & \\ \hline
\end{tabular}
\caption{Matter spectrum obtained after switching on the WLs, the numbers in parenthesis label its corresponding representation under $SU(3)_C\times SU(2)_L\times SU(3)\times SU(2)$. The subindex labels the hypercharge.}
\label{tab:SpectrumSM}
\end{center}
\end{table}
\subsection{Prospects: VEVs, Light Higgses and Decoupled Exotics}
\label{muproblem}
In order to retain a reasonable phenomenology we have to give VEVs to some singlet fields. The VEV configuration is expected to cancel the FI term induced by the anomalous $U(1)$ and remove dangerous exotics from the low energy spectrum by means of the Froggatt-Nielsen mechanism. Depending on the VEV fields, the $U(1)$s as well as discrete $R$ and non-$R$ symmetries will be generically broken in such a way that certain discrete combinations survive. Those combinations are of particular interest for us, since they can be used to control some dangerous operators and in particular, can be used to set the $\mu$ term to vanish in the supersymmetric vacuum.

In principle, the choice of the VEVs is not entirely arbitrary since one has to ensure $F$- and $D$-flatness of the new vacuum, however, given the large amount of VEVs and their generic size, it is assumed that such flat directions exist. Our main goal is to ensure the absence of the $\mu$ term in the superpotential, the common consensus is that this happens due to an $R$-symmetry. In the context of the \ztt\ orbifold \cite{Blaszczyk:2009in}, some surviving $R$-symmetries in the low energy have been engineered \cite{Kappl:2010yu}.  Here we try the simplest alternatives for an $R$-symmetry, which turn out to have significant drawbacks, but we expect that in the more comprehensive exploration we are currently performing, such drawbacks can be avoided.

The \ztf\ orbifold allows for three $\mathbbm{Z}_2$ $R$-symmetries one per each complex plane, given an $L$ point coupling $\Phi_1\cdots \Phi_L$, it is only allowed if it fulfills the conditions
\begin{align*}
\sum_{k=1}^L R_k^{(i)}=-1\,\, {\rm mod}\,\, 2\, , \quad i=1,2,3\, ,
\end{align*}
in which $R_k^{(i)}$ denotes the $R$-charge of the $k$-th field and it is given by \cite{Kobayashi:2004ya}
\begin{align*}
R_k^{(i)}=q^i_k - N^i_k + \tilde{N}^i_k\, ,
\end{align*}
where the weight $q^i$ encodes the information about the fermionic excitations (see table \ref{tab:rmovers}) and $N^i$ ($\tilde{N}^i$) counts the number of (anti-) holomorphic bosonic oscillators used to construct the physical state. With this information in mind we can reconsider the GUT bilinear
\begin{align*}
( \mathbf{10},  \mathbf{2},  \mathbf{2},  \mathbf{1})_{ 0,    0} \cdot ( \mathbf{10},  \mathbf{2},  \mathbf{2},  \mathbf{1})_{ 0,    0}\supset H_u H_d\, .
\end{align*}
This term is neutral under all selection rules, so that the coupling $\mu H_u H_d$ is only absent in the superpotential if there is a leftover $R$-symmetry after the VEV configuration is chosen. From table \ref{tab:rmovers} we can see that the $R$-charges are in close relation with the twisted sector. In particular note that by taking only VEV fields from $T_{(0,1)}$, $T_{(0,2)}$, $T_{(0,3)}$ and untwisted fields with $q=(0,0,-1,0)$ or $q=(0,0,0,-1)$, the $\mathbbm{Z}_2$ $R$-symmetry from the first torus survives. However this did not work out because after we gave VEVs to all possible singlets in the mentioned sectors, two $U(1)$s remain unbroken, and some SM fields are charged under them.
\begin{table}[h!]
\renewcommand{\arraystretch}{1.15}
\begin{center}
\begin{tabular}{c c}
\begin{tabular}{|l|l|} \hline
\multirow {3}*{$U$} & $(0,-1,0,0)$\\
    &  $(0,0,-1,0)$\\
  & $(0,0,0,-1)$\\ \hline
$T_{(0,1)}$  &  $-\frac14 (0,0,1,3)$\\ \hline
$T_{(0,2)}$  &  $-\frac12 (0,0,1,1)$\\ \hline
\end{tabular}
&
\begin{tabular}{|l|l|}\hline
$T_{(0,3)}$  &  $-\frac14 (0,0,3,1)$\\ \hline
$T_{(1,0)}$  & $-\frac12 (0,1,0,1)$\\ \hline
$T_{(1,1)}$  & \multicolumn{1}{c|}{-}\\ \hline
$T_{(1,2)}$  & $-\frac12 (0,1,1,0)$\\ \hline
$T_{(1,3)}$  & $-\frac14 (0,2,1,1)$\\ \hline
\end{tabular}
\end{tabular}
\caption{Right moving weights $q$ for the various sectors of \ztf.}
\label{tab:rmovers}
\end{center}
\end{table}\\
Note from table \ref{tab:rmovers} that all spinor weights add up to $-1$, so that if one takes the sum of the three $R$-symmetries
\begin{align}
\sum_{k=1}^L \left(\sum_{i=1}^3R_k^{(i)}\right)=-1\,\, {\rm mod}\,\, 2\, \label{sum}
\end{align}
states with no oscillators transform with $R$-charge $-1$, whereas oscillator states have charges $0,-2$. Hence, if we take VEVs of oscillator states, the previous symmetry survives. There are enough oscillator singlets to break all $U(1)$ symmetries. The drawback of this symmetry relies on the charge assignment for the fields and the constraints it imposes on the mass terms. Take for instance a non-oscillator field $\Phi$, if the $R$-symmetry in \eqref{sum} survives, $\Phi$ can only get a mass term when coupled to a field $\Phi^\prime$ which contains oscillators. Due to the previous argument many exotics remain massless after we consider the VEV configuration.

We have thus observed that a surviving $R$-symmetry which alleviates the $\mu$ problem is an alternative if the following two conditions are satisfied
\begin{itemize}
\item There are enough singlet fields with $R$-charge zero, so that their corresponding VEVs suffice the breaking of all extra $U(1)$s.
\item The $R$-charge assignment for the fields allows for a mass term for all exotics in the model. Though this constraint is very hard to satisfy, one can look for situations in which the surviving $R$-symmetry is a mixture of any of the original ones with some non-$R$ factors. There one expects more intricate charge assignments and hence more mass terms to be allowed, in contrast to the previous example.
\end{itemize}
\section{Conclusions}
\label{con}
In this work we wanted to describe general results from model building on the recently constructed \ztf\ orbifold. The localization of the SM fields in the extra dimensions follows the general pattern identified in the \zsix\ Mini--Landscape, and we expect these locations to play
a crucial role in solving field theory issues such as the doublet-triplet splitting,
flavor and $\mu$- problems. The lessons we learned from the geography can be summarized as follows
\begin{enumerate}
\item A scenario in which the three families arise from complete
\SO{10} multiplets is not consistent with a hierarchy for the mass of
the SM fields.
\item A completely untwisted top-Yukawa coupling seems to be the most
favored situation, leading to the familiar gauge-top unification
scheme.
\item In most of the cases the down-type Higgs lives
in the bulk as well. If the untwisted Higgs pair remains massless the model will enjoy of gauge-Higgs unification.
\item The two light families usually arise from the twisted sectors.
They can appear as complete multiplets of the underlying local GUT.
\end{enumerate}
It is worth to remark that these conclusions were drawn from models in which a GUT factor ($SO(10)$ or $E_6$) is retained at the stage in which all Wilson lines are off. We also wanted to show that this was a viable strategy by searching for explicit Wilson line configurations. Taking one of the shift embeddings from ref. \cite{Pena:2012ki} we constructed MI Wilson lines aided by the C++ \mbox{Orbifolder}, the search led us to around two hundred inequivalent configurations. We have then shown that the top-Yukawa at trilinear order can survive the Wilson line projections and that extra matter fields arising from unshielded singularities either complete the top family or contribute with vector-like exotics. One step further, we tried certain VEV configurations in the spirit of keeping one $R$-symmetry unbroken. As an outcome we found that these configurations usually leave many exotic fields massless. It is thus a matter under exploration whether or not this model allows for an $R$-symmetry which could forbid the $\mu$ term.

\acknowledgments

We would like to thank Hans Peter Nilles, with whom most of this work was done. We also thank Michael Blaszczyk, Nana Cabo Bizet, Sven Krippendorf, Matthias Schmitz and Fabian R\"uhle for helpful
discussions and suggestions. This work was partially supported by the
SFB-Transregio TR33 The Dark Universe (Deutsche Forschungsgemeinschaft) and the European Union
7th network program Unification in the LHC era (PITN-GA-2009-237920). We thank the
Simons Center for Geometry and Physics in Stony Brook (NY) for
hospitality during the completion of this work.
\appendix
\section{Promising $E_6$ Embeddings}
\label{e6}
Here we present a selection of $E_6$ models with a Wilson line configuration leading to two complete families ($\mathbf{27}$) and an untwisted trilinear coupling. All of these models work with all Wilson lines on except $W_1$ or $W_2$.

For the shift embedding
\begin{align*}
4V_4=&(0,0,0,0,0,0,0,0)(1,1,0,0,0,0,0,0)\, ,\\
2V_2=&(-1,-1,-1,0,0,0,0,1)(0,1,0,0,0,0,1,0) \, ,
\end{align*}
if we take $W_{1,2}\neq 0$ to break $E_6$ to $SO(10)$ the surviving pieces from the split multiplets may lead to complete families, provided the $\mathbf{16}{\rm -plets}\subset \mathbf{27}$ are not projected out. In the case of
\begin{align*}
4V_4=&(2,2,0,0,0,0,0,0)(1,1,0,0,0,0,0,0)\, , \\
2V_2=&(2,2,0,0,0,0,0,0)(0,1,0,0,0,0,1,0) \,
\end{align*}
we have to switch on the WLs in the same manner as before, taking again $W_{1,2}$ to break to $SO(10)$; Then we have the chance to obtain two $\mathbf{16}$-plets at the $T_{(1,2)}$ sector. The model given by
\begin{align*}
4V_4=&(1,1,0,0,0,0,0,0)(1,1,1,1,1,1,1,-1)\,,\\
2V_2=&(0,1,0,0,0,0,1,0)(1,1,1,1,1,1,1,-1)\,,
\end{align*}
may lead to two families at the $T_{(1,2)}$ sector, provided $W_3$ breaks to $SO(10)$. This picture is similar to that one gets from the embedding
\begin{align*}
4V_4=&(1,1,0,0,0,0,0,0)(1,1,1,1,1,1,1,-1)\,,\\
2V_2=&(0,1,0,0,0,0,1,0)(2,0,0,0,0,-2,0,0)\,,
\end{align*}
where the only difference is that the families will appear at the $T_{(1,0)}$ sector.

\section{Complete Spectrum for the Benchmark Model}
\label{app:Spectrum}
Here we will give the complete spectrum for the model discussed in sect. \ref{explicit}. The first U(1) is anomalous and the gauge group is
given by $SU(3) \times SU(2) \times SU(3) \times SU(2) \times U(1)^9$\\

Untwisted Sector:
\begin{longtable}{|r|c|l|}
\hline
 \# & irrep & $\text{U}(1)$ charges\\
\hline\hline
\endhead
\hline
\endfoot
    1 & $\left(\crep{3},\rep{1},\rep{1},\rep{1}\right)_{l}$ &
$\left(\tfrac{5}{4},   -5,  129,  214,    0,    9,   -9,  -27,   18,
34\right)$\\
    1 & $\left(\rep{1},\rep{1},\rep{1},\rep{1}\right)_{l}$ &
$\left(\tfrac{5}{4},   -5,  129, -192,    0,   13,  -13,  -39,   26,
-122\right)$\\
    1 & $\left(\rep{1},\rep{1},\rep{1},\rep{1}\right)_{l}$ &
$\left(\tfrac{1}{4},    0,    0,    0,    2,  205, -136,  152, -227,
-1\right)$\\
    1 & $\left(\rep{1},\rep{1},\crep{3},\rep{1}\right)_{l}$ &
$\left(\tfrac{5}{4},    0,    0,    0,    0, -205,   -2,  274,  -57,
-5\right)$\\
    1 & $\left(\rep{1},\rep{2},\rep{1},\rep{1}\right)_{l}$ & $\left(
-3,   -3,  -65, -229,    0,  -24,   24,   72,  -48,   12\right)$\\
    1 & $\left(\rep{1},\rep{2},\rep{1},\rep{1}\right)_{l}$ & $\left(
3,    3,   65,  229,    0,   24,  -24,  -72,   48,  -12\right)$\\
    1 & $\left(\rep{1},\rep{1},\rep{1},\rep{2}\right)_{l}$ &
$\left(\tfrac{3}{4},    0,    0,    0,   -1, -205, -140,  140, -219,
-3\right)$\\
    1 & $\left(\rep{3},\rep{2},\rep{1},\rep{1}\right)_{l}$ &
$\left(\tfrac{7}{4},    8,  -64,   15,    0,   15,  -15,  -45,   30,
-46\right)$\\
\end{longtable}

$T_{(0,1)}$ Sector:
\begin{longtable}{|r|c|l|}
\hline
 \# & irrep & $\text{U}(1)$ charges\\
\hline\hline
\endhead
\hline
\endfoot
    1 & $\left(\rep{1},\rep{1},\rep{3},\rep{1}\right)_{l}$ &
$\left(\tfrac{3}{8},\tfrac{13}{2},-\tfrac{15}{2},   -3,
0,\tfrac{199}{2},    4,  222,
-148,-\tfrac{3}{2}\right)$\\

    1 & $\left(\rep{1},\rep{1},\rep{1},\rep{1}\right)_{l}$ &
$\left(\tfrac{7}{8},\tfrac{13}{2},-\tfrac{15}{2},   -3,
0,-\tfrac{211}{2},  209,  137,
-217,-\tfrac{7}{2}\right)$\\

    1 & $\left(\rep{1},\rep{1},\rep{1},\rep{2}\right)_{l}$ &
$\left(\tfrac{3}{8},\tfrac{13}{2},-\tfrac{15}{2},   -3,
1,-\tfrac{211}{2},  -67,  149,
-225,-\tfrac{3}{2}\right)$\\

    1 & $\left(\rep{1},\rep{1},\rep{1},\rep{1}\right)_{l}$ &
$\left(-\tfrac{3}{8},\tfrac{13}{2},-\tfrac{15}{2},   -3,
2,\tfrac{199}{2},   73,    9,
-6,\tfrac{3}{2}\right)$\\

    1 & $\left(\rep{1},\rep{1},\rep{1},\rep{1}\right)_{l}$ &
$\left(\tfrac{1}{8},\tfrac{13}{2},-\tfrac{15}{2},   -3,
0,-\tfrac{211}{2},    2,  216,
233,-\tfrac{1}{2}\right)$\\

    1 & $\left(\rep{1},\rep{1},\rep{1},\rep{1}\right)_{l}$ &
$\left(-\tfrac{1}{8},   -4, -146,  104,    0,  104, -104,  -32, -230,
-58\right)$\\
    1 & $\left(\rep{1},\rep{1},\rep{1},\rep{1}\right)_{l}$ &
$\left(\tfrac{23}{8},   -1,   97,  120,    0,  -85,   85,  115,   49,
-70\right)$\\
    1 & $\left(\rep{1},\rep{1},\rep{3},\rep{1}\right)_{l}$ &
$\left(\tfrac{7}{8},    4,  -32,  -94,    0,  111,   96, -132,   88,
55\right)$\\
    1 & $\left(\rep{1},\rep{1},\rep{1},\rep{2}\right)_{l}$ &
$\left(\tfrac{7}{8},    4,  -32,  -94,    1,  -94,   25, -205,   11,
55\right)$\\
    1 & $\left(\rep{1},\rep{1},\rep{1},\rep{1}\right)_{l}$ &
$\left(\tfrac{17}{8},    4,  -32,  -94,    0,  -94,   94,  142,   31,
50\right)$\\
    1 & $\left(\crep{3},\rep{1},\rep{1},\rep{1}\right)_{l}$ &
$\left(-\tfrac{1}{8},    0,    0,    0,    0,-\tfrac{205}{2},   -1,
-213, -235,\tfrac{79}{2}\right)$\\
    1 & $\left(\rep{1},\rep{1},\rep{1},\rep{1}\right)_{l}$ &
$\left(-\tfrac{5}{8},    0,    0,    0,    0,-\tfrac{205}{2},   -1,
-213, -235,-\tfrac{229}{2}\right)$\\
    1 & $\left(\rep{1},\rep{1},\rep{1},\rep{2}\right)_{l}$ &
$\left(\tfrac{11}{8},-\tfrac{5}{2},-\tfrac{49}{2},  -91,    1,  114,
24,  -68, -206,  -64\right)$\\
    1 & $\left(\crep{3},\rep{1},\rep{1},\rep{1}\right)_{l}$ &
$\left(-\tfrac{13}{8},\tfrac{5}{2},\tfrac{49}{2},   91,    0,   91,
116,   68,  206,  -13\right)$\\
    1 & $\left(\rep{1},\rep{1},\rep{1},\rep{2}\right)_{l}$ &
$\left(\tfrac{15}{8},-\tfrac{5}{2},-\tfrac{49}{2},  -91,   -1,  -91,
-47,  139,   33,  -66\right)$\\
    1 & $\left(\rep{1},\rep{1},\rep{1},\rep{1}\right)_{l}$ &
$\left(-\tfrac{5}{8},\tfrac{5}{2},-\tfrac{307}{2},  101,    0, -104,
-103, -169,  -13,   61\right)$\\
    1 & $\left(\rep{1},\rep{1},\rep{1},\rep{1}\right)_{l}$ &
$\left(\tfrac{5}{8},-\tfrac{5}{2},-\tfrac{49}{2},  -91,    0,  -91,
-116, -208,   13,  -61\right)$\\
    1 & $\left(\rep{1},\rep{1},\rep{1},\rep{1}\right)_{l}$ &
$\left(\tfrac{9}{8},-\tfrac{5}{2},-\tfrac{49}{2},  -91,    0,  114,
93,   -1,  252,  -63\right)$\\
\end{longtable}

$T_{(0,2)}$ Sector:
\begin{longtable}{|r|c|l|}
\hline
 \# & irrep & $\text{U}(1)$ charges\\
\hline\hline
\endhead
\hline
\endfoot
    4 & $\left(\rep{3},\rep{1},\rep{1},\rep{1}\right)_{l}$ &
$\left(-\tfrac{1}{2},    0,    0,    0,    0,    0,    0, -140,  219,
80\right)$\\
    2 & $\left(\rep{1},\rep{1},\rep{1},\rep{1}\right)_{l}$ & $\left(
2,    0, -178,   10,    0,   10,  -10,  110, -199,   -8\right)$\\
    2 & $\left(\crep{3},\rep{1},\rep{1},\rep{1}\right)_{l}$ & $\left(
-1,    0,    0,    0,    0,    0,    0, -140,  219,  -74\right)$\\
    2 & $\left(\rep{1},\rep{1},\rep{1},\rep{1}\right)_{l}$ &
$\left(-\tfrac{1}{2},    0,  178,  -10,    0,  -10,   10,  170, -239,
2\right)$\\
    2 & $\left(\crep{3},\rep{1},\rep{1},\rep{1}\right)_{l}$ &
$\left(\tfrac{7}{4},-\tfrac{5}{2},-\tfrac{49}{2},  -91,
0,\tfrac{23}{2}, -115,  145,
29,-\tfrac{53}{2}\right)$\\

    2 & $\left(\rep{1},\rep{1},\rep{1},\rep{1}\right)_{l}$ & $\left(
0,-\tfrac{5}{2},\tfrac{307}{2}, -101,    0,\tfrac{3}{2},  102, -184,
-3,\tfrac{117}{2}\right)$\\

    2 & $\left(\rep{3},\rep{1},\rep{1},\rep{1}\right)_{l}$ & $\left(
-1,\tfrac{5}{2},\tfrac{49}{2},   91,    0,-\tfrac{23}{2},  -92,  214,
-17,\tfrac{47}{2}\right)$\\
    2 & $\left(\rep{1},\rep{1},\rep{1},\rep{1}\right)_{l}$ &
$\left(-\tfrac{3}{4},\tfrac{5}{2},-\tfrac{307}{2},  101,
0,-\tfrac{3}{2},  105, -175,
-9,-\tfrac{111}{2}\right)$\\

    2 & $\left(\rep{3},\rep{1},\rep{1},\rep{1}\right)_{l}$ & $\left(
0,-\tfrac{13}{2},\tfrac{15}{2},    3,    0,  208,   -1,   -3,    2,
-39\right)$\\
    2 & $\left(\rep{1},\rep{1},\rep{1},\rep{1}\right)_{l}$ &
$\left(\tfrac{1}{2},-\tfrac{13}{2},\tfrac{15}{2},    3,    0,  208,
-1,   -3,    2,  115\right)$\\
    2 & $\left(\rep{1},\rep{2},\rep{1},\rep{1}\right)_{l}$ &
$\left(\tfrac{3}{4},\tfrac{3}{2},\tfrac{243}{2},    8,    0,  213,
-6,  -18,   12,   -3\right)$\\
    2 & $\left(\rep{1},\rep{1},\rep{1},\rep{1}\right)_{l}$ &
$\left(\tfrac{9}{4},\tfrac{3}{2},-\tfrac{113}{2},  221,    0, -189,
-18,  -54,   36,   -9\right)$\\
    2 & $\left(\rep{1},\rep{1},\rep{1},\rep{1}\right)_{l}$ &
$\left(\tfrac{11}{4},\tfrac{3}{2},-\tfrac{113}{2}, -185,    0, -185,
-22,  -66,   44,  -11\right)$\\
    2 & $\left(\rep{1},\rep{1},\crep{3},\rep{1}\right)_{l}$ & $\left(
-1,   -4,   32,   94,    0,-\tfrac{17}{2},  112,  -14,
135,\tfrac{125}{2}\right)$\\
    2 & $\left(\rep{1},\rep{1},\rep{1},\rep{1}\right)_{l}$ &
$\left(-\tfrac{3}{4},   -4,   32,   94,    0,\tfrac{393}{2},  114,   -8,
-246,\tfrac{123}{2}\right)$\\
    2 & $\left(\rep{1},\rep{1},\rep{1},\rep{2}\right)_{l}$ &
$\left(\tfrac{3}{2},    4,  -32,  -94,    1,\tfrac{17}{2},   26,  148,
27,-\tfrac{129}{2}\right)$\\
    2 & $\left(\rep{1},\rep{1},\rep{1},\rep{1}\right)_{l}$ &
$\left(\tfrac{1}{4},    4,  -32,  -94,    2,\tfrac{17}{2},  -43,
-199,    7,-\tfrac{119}{2}\right)$\\
    2 & $\left(\rep{1},\rep{1},\rep{1},\rep{1}\right)_{l}$ &
$\left(\tfrac{3}{2},    4,  -32,  -94,    0,-\tfrac{393}{2},   93,  -71,
-204,-\tfrac{129}{2}\right)$\\
\end{longtable}

$T_{(0,3)}$ Sector:
\begin{longtable}{|r|c|l|}
\hline
 \# & irrep & $\text{U}(1)$ charges\\
\hline\hline
\endhead
\hline
\endfoot
    1 & $\left(\rep{1},\rep{1},\rep{1},\rep{2}\right)_{l}$ &
$\left(\tfrac{9}{8},-\tfrac{13}{2},\tfrac{15}{2},    3,
-1,\tfrac{211}{2},   67,  131,
-213,-\tfrac{9}{2}\right)$\\

    1 & $\left(\rep{1},\rep{1},\crep{3},\rep{1}\right)_{l}$ &
$\left(\tfrac{9}{8},-\tfrac{13}{2},\tfrac{15}{2},    3,
0,-\tfrac{199}{2},   -4,   58,
-290,-\tfrac{9}{2}\right)$\\

    1 & $\left(\rep{1},\rep{1},\rep{1},\rep{1}\right)_{l}$ &
$\left(\tfrac{5}{8},-\tfrac{13}{2},\tfrac{15}{2},    3,
0,\tfrac{211}{2}, -209,  143,
-221,-\tfrac{5}{2}\right)$\\

    1 & $\left(\rep{1},\rep{1},\rep{1},\rep{1}\right)_{l}$ &
$\left(-\tfrac{1}{8},-\tfrac{13}{2},\tfrac{15}{2},    3,
0,\tfrac{211}{2},   -2, -216,
-233,\tfrac{1}{2}\right)$\\

    1 & $\left(\rep{1},\rep{1},\rep{1},\rep{1}\right)_{l}$ &
$\left(\tfrac{3}{8},-\tfrac{13}{2},\tfrac{15}{2},    3,
-2,-\tfrac{199}{2},  -73,   -9,
6,-\tfrac{3}{2}\right)$\\

    1 & $\left(\rep{3},\rep{1},\rep{1},\rep{2}\right)_{l}$ &
$\left(\tfrac{1}{8},    0,    0,    0,   -1,-\tfrac{205}{2},   68,
-6,    4,-\tfrac{79}{2}\right)$\\
    1 & $\left(\rep{1},\rep{1},\rep{1},\rep{1}\right)_{l}$ &
$\left(\tfrac{1}{8},    0,    0,    0,    0,-\tfrac{205}{2}, -208,
6,   -4,\tfrac{233}{2}\right)$\\
    1 & $\left(\rep{3},\rep{1},\rep{1},\rep{1}\right)_{l}$ &
$\left(-\tfrac{3}{8},    0,    0,    0,    0,-\tfrac{205}{2}, -208,
6,   -4,-\tfrac{75}{2}\right)$\\
    1 & $\left(\rep{1},\rep{1},\rep{1},\rep{2}\right)_{l}$ &
$\left(\tfrac{5}{8},    0,    0,    0,   -1,-\tfrac{205}{2},   68,
-6,    4,\tfrac{229}{2}\right)$\\
    1 & $\left(\rep{1},\rep{1},\rep{1},\rep{1}\right)_{l}$ &
$\left(\tfrac{9}{8},   -9,  -17,  -88,    0,  117, -117, -211,   15,
54\right)$\\
    1 & $\left(\rep{1},\rep{1},\rep{1},\rep{1}\right)_{l}$ &
$\left(-\tfrac{7}{8},   -4,   32,   94,   -2, -111,   42,  -14, -242,
-55\right)$\\
    1 & $\left(\rep{1},\rep{2},\rep{1},\rep{1}\right)_{l}$ &
$\left(\tfrac{1}{8},    4,  -32,  109,    0,  109, -109, -187,   -1,
58\right)$\\
    1 & $\left(\crep{3},\rep{1},\rep{1},\rep{1}\right)_{l}$ &
$\left(\tfrac{9}{8},    4,  -32,  -94,    0,  -94,   94,    2,  250,
-24\right)$\\
    1 & $\left(\rep{1},\rep{1},\rep{1},\rep{1}\right)_{l}$ &
$\left(-\tfrac{17}{8},   -4,   32,   94,    0,   94,  -94, -142,  -31,
-50\right)$\\
    1 & $\left(\rep{3},\rep{1},\rep{1},\rep{1}\right)_{l}$ &
$\left(\tfrac{17}{8},-\tfrac{5}{2},-\tfrac{49}{2},  -91,    0,  114,
93,  139,   33,   11\right)$\\
    1 & $\left(\rep{1},\rep{1},\rep{1},\rep{1}\right)_{l}$ &
$\left(\tfrac{1}{8},-\tfrac{5}{2},\tfrac{307}{2}, -101,    0, -101,
-106,  -38, -226,  -59\right)$\\
    1 & $\left(\rep{1},\rep{1},\rep{1},\rep{1}\right)_{l}$ &
$\left(-\tfrac{9}{8},\tfrac{5}{2},\tfrac{49}{2},   91,    0, -114,
-93,    1, -252,   63\right)$\\
    1 & $\left(\rep{1},\rep{1},\rep{1},\rep{1}\right)_{l}$ &
$\left(-\tfrac{5}{8},\tfrac{5}{2},\tfrac{49}{2},   91,    0,   91,
116,  208,  -13,   61\right)$\\
\end{longtable}

$T_{(1,0)}$ Sector:
\begin{longtable}{|r|c|l|}
\hline
 \# & irrep & $\text{U}(1)$ charges\\
\hline\hline
\endhead
\hline
\endfoot
    2 & $\left(\rep{1},\rep{1},\rep{1},\rep{1}\right)_{l}$ &
$\left(-\tfrac{5}{8},   -4,   32, -109,    2,   96,  -27,
129,\tfrac{205}{2},   61\right)$\\
    2 & $\left(\rep{1},\rep{1},\rep{1},\rep{2}\right)_{l}$ &
$\left(-\tfrac{5}{8},   -4,   32, -109,   -1,   96,  -27,
129,\tfrac{205}{2},   61\right)$\\
    2 & $\left(\rep{1},\rep{1},\rep{1},\rep{1}\right)_{l}$ &
$\left(\tfrac{5}{8},   -4,   32, -109,    0, -109,  109,
257,-\tfrac{217}{2},   56\right)$\\
    2 & $\left(\rep{1},\rep{1},\crep{3},\rep{1}\right)_{l}$ &
$\left(\tfrac{5}{8},    4,  -32,  109,    0,  -96, -111,
17,\tfrac{103}{2},  -61\right)$\\
    2 & $\left(\rep{1},\rep{1},\rep{1},\rep{1}\right)_{l}$ &
$\left(\tfrac{7}{8},    4,  -32,  109,    0,  109, -109,
23,-\tfrac{659}{2},  -62\right)$\\
    2 & $\left(\rep{1},\rep{1},\rep{1},\rep{1}\right)_{l}$ &
$\left(-\tfrac{25}{8},-\tfrac{3}{2},\tfrac{113}{2},  -18,
0,\tfrac{169}{2},   19,
-223,-\tfrac{331}{2},\tfrac{25}{2}\right)$\\

    2 & $\left(\rep{1},\rep{1},\rep{3},\rep{1}\right)_{l}$ &
$\left(\tfrac{15}{8},\tfrac{3}{2},-\tfrac{113}{2},   18,
0,\tfrac{241}{2},  -17,
-51,\tfrac{445}{2},-\tfrac{15}{2}\right)$\\

    4 & $\left(\crep{3},\rep{1},\rep{1},\rep{1}\right)_{l}$ &
$\left(\tfrac{15}{8},\tfrac{3}{2},\tfrac{65}{2},   13,   -1,
13,-\tfrac{95}{2},-\tfrac{145}{2},
-203,\tfrac{63}{2}\right)$\\

    2 & $\left(\rep{1},\rep{1},\rep{1},\rep{1}\right)_{l}$ &
$\left(\tfrac{11}{8},\tfrac{3}{2},\tfrac{65}{2},   13,   -1,
13,-\tfrac{95}{2},-\tfrac{145}{2},
-203,-\tfrac{245}{2}\right)$\\

    2 & $\left(\rep{1},\rep{1},\rep{1},\rep{1}\right)_{l}$ &
$\left(\tfrac{7}{8},   -9,   72,  110,
-1,\tfrac{15}{2},-\tfrac{291}{2},-\tfrac{33}{2},   11,
55\right)$\\

    2 & $\left(\rep{3},\rep{1},\rep{1},\rep{1}\right)_{l}$ &
$\left(\tfrac{1}{8},   -4,  -57, -104,
1,-\tfrac{3}{2},\tfrac{279}{2},-\tfrac{3}{2},    1,   19\right)$\\
    2 & $\left(\crep{3},\rep{1},\rep{1},\rep{1}\right)_{l}$ &
$\left(\tfrac{3}{8},    4,   57,  104,
1,\tfrac{3}{2},\tfrac{273}{2},-\tfrac{21}{2},    7,
-21\right)$\\

    2 & $\left(\rep{1},\rep{1},\rep{1},\rep{1}\right)_{l}$ &
$\left(-\tfrac{11}{8},    9,  -72, -110,
-1,-\tfrac{15}{2},-\tfrac{261}{2},\tfrac{57}{2},  -19,
-53\right)$\\

    2 & $\left(\rep{1},\rep{2},\rep{1},\rep{1}\right)_{l}$ &
$\left(\tfrac{7}{8},-\tfrac{13}{2},-\tfrac{163}{2},    8,    1,
8,\tfrac{53}{2},\tfrac{19}{2},
245,-\tfrac{7}{2}\right)$\\

    2 & $\left(\rep{1},\rep{1},\rep{1},\rep{1}\right)_{l}$ &
$\left(-\tfrac{1}{8},-\tfrac{13}{2},\tfrac{193}{2}, -205,    1,
0,\tfrac{69}{2},\tfrac{67}{2},
229,\tfrac{1}{2}\right)$\\

\end{longtable}

$T_{(1,1)}$ Sector: empty\\

$T_{(1,2)}$ Sector:
\begin{longtable}{|r|c|l|}
\hline
 \# & irrep & $\text{U}(1)$ charges\\
\hline\hline
\endhead
\hline
\endfoot
    2 & $\left(\rep{1},\rep{1},\rep{1},\rep{1}\right)_{l}$ &
$\left(\tfrac{3}{8},    0,    0,  203,    0,\tfrac{201}{2},    3,
289,\tfrac{243}{2},-\tfrac{3}{2}\right)$\\

    2 & $\left(\rep{1},\rep{1},\rep{1},\rep{2}\right)_{l}$ &
$\left(-\tfrac{7}{8},    0,    0,  203,    1,\tfrac{201}{2},  -66,
-58,\tfrac{203}{2},\tfrac{7}{2}\right)$\\

    2 & $\left(\rep{1},\rep{1},\rep{1},\rep{1}\right)_{l}$ &
$\left(\tfrac{1}{8},    0,    0, -203,    0,\tfrac{209}{2},  206,
-82,\tfrac{235}{2},-\tfrac{1}{2}\right)$\\

    2 & $\left(\rep{1},\rep{1},\rep{1},\rep{1}\right)_{l}$ &
$\left(-\tfrac{7}{8},    0,    0,  203,   -2,\tfrac{201}{2},  -66,
-58,\tfrac{203}{2},\tfrac{7}{2}\right)$\\

    2 & $\left(\rep{1},\rep{1},\crep{3},\rep{1}\right)_{l}$ &
$\left(\tfrac{3}{8},    0,    0, -203,    0,\tfrac{209}{2},   -1,
-3,-\tfrac{373}{2},-\tfrac{3}{2}\right)$\\

    2 & $\left(\rep{1},\rep{1},\rep{1},\rep{1}\right)_{l}$ &
$\left(-\tfrac{7}{8},-\tfrac{3}{2},-\tfrac{243}{2},   -8,
0,-\tfrac{221}{2},    7,
161,\tfrac{665}{2},\tfrac{7}{2}\right)$\\

    2 & $\left(\rep{1},\rep{1},\rep{1},\rep{1}\right)_{l}$ &
$\left(\tfrac{9}{8},\tfrac{3}{2},\tfrac{243}{2},    8,
0,-\tfrac{189}{2},   -9,
113,\tfrac{729}{2},-\tfrac{9}{2}\right)$\\

    4 & $\left(\rep{1},\rep{1},\rep{1},\rep{1}\right)_{l}$ &
$\left(\tfrac{5}{8},-\tfrac{13}{2},-\tfrac{163}{2}, -195,   -1,
10,-\tfrac{89}{2},-\tfrac{407}{2},
10,-\tfrac{5}{2}\right)$\\

    4 & $\left(\rep{1},\rep{1},\rep{1},\rep{1}\right)_{l}$ &
$\left(-\tfrac{11}{8},-\tfrac{3}{2},-\tfrac{65}{2},  -13,    1,
-13,\tfrac{95}{2},\tfrac{425}{2},
-16,-\tfrac{223}{2}\right)$\\

    2 & $\left(\rep{1},\rep{1},\rep{1},\rep{1}\right)_{l}$ &
$\left(\tfrac{21}{8},\tfrac{3}{2},\tfrac{65}{2},   13,    1,
13,\tfrac{43}{2},\tfrac{269}{2},
36,\tfrac{213}{2}\right)$\\

    2 & $\left(\rep{1},\rep{1},\rep{3},\rep{1}\right)_{l}$ &
$\left(\tfrac{19}{8},   -5,   40,   16,    1,
16,\tfrac{37}{2},-\tfrac{29}{2},
-116,-\tfrac{19}{2}\right)$\\

    2 & $\left(\rep{1},\rep{1},\rep{1},\rep{1}\right)_{l}$ &
$\left(\tfrac{23}{8},   -5,   40,   16,   -1,
16,\tfrac{313}{2},\tfrac{239}{2},
46,-\tfrac{23}{2}\right)$\\

    2 & $\left(\rep{1},\rep{1},\rep{1},\rep{2}\right)_{l}$ &
$\left(\tfrac{13}{8},   -5,   40,   16,    0,
16,\tfrac{175}{2},-\tfrac{455}{2},
26,-\tfrac{13}{2}\right)$\\

    2 & $\left(\rep{1},\rep{1},\rep{1},\rep{1}\right)_{l}$ &
$\left(-\tfrac{19}{8},    5,  -40,  -16,    1,
189,\tfrac{105}{2},\tfrac{175}{2},
193,\tfrac{19}{2}\right)$\\

    2 & $\left(\rep{1},\rep{1},\rep{1},\rep{1}\right)_{l}$ &
$\left(-\tfrac{15}{8},    5,  -40,  -16,    1,
-221,\tfrac{97}{2},\tfrac{151}{2},
201,\tfrac{15}{2}\right)$\\

    2 & $\left(\rep{1},\rep{1},\rep{1},\rep{1}\right)_{l}$ &
$\left(-\tfrac{15}{8},\tfrac{13}{2},\tfrac{163}{2},  195,   -1,
-10,-\tfrac{49}{2},-\tfrac{287}{2},
-30,\tfrac{15}{2}\right)$\\

\end{longtable}

$T_{(1,3)}$ Sector:
\begin{longtable}{|r|c|l|}
\hline
 \# & irrep & $\text{U}(1)$ charges\\
\hline\hline
\endhead
\hline
\endfoot
    2 & $\left(\rep{3},\rep{2},\rep{1},\rep{1}\right)_{l}$ &
$\left(\tfrac{1}{4},    0,    0,    0,    0,    0,    0,
70,-\tfrac{219}{2},  -40\right)$\\
    2 & $\left(\rep{1},\rep{2},\rep{1},\rep{1}\right)_{l}$ &
$\left(\tfrac{3}{4},    0,    0,    0,    0,    0,    0,
70,-\tfrac{219}{2},  114\right)$\\
    2 & $\left(\crep{3},\rep{1},\rep{1},\rep{1}\right)_{l}$ &
$\left(\tfrac{1}{4},    0,    0,  203,    0,   -2,    2,
76,-\tfrac{227}{2},   38\right)$\\
    2 & $\left(\rep{1},\rep{1},\rep{1},\rep{1}\right)_{l}$ &
$\left(\tfrac{1}{4},    0,    0, -203,    0,    2,   -2,
64,-\tfrac{211}{2}, -118\right)$\\
    2 & $\left(\rep{1},\rep{1},\rep{1},\rep{1}\right)_{l}$ &
$\left(-\tfrac{1}{4},    0,    0,  203,    0,   -2,    2,
76,-\tfrac{227}{2}, -116\right)$\\
    2 & $\left(\crep{3},\rep{1},\rep{1},\rep{1}\right)_{l}$ &
$\left(\tfrac{3}{4},    0,    0, -203,    0,    2,   -2,
64,-\tfrac{211}{2},   36\right)$\\
    2 & $\left(\rep{1},\rep{1},\rep{1},\rep{1}\right)_{l}$ & $\left(
-1,    5, -129,  -11,    0,  -11,   11,  103,-\tfrac{263}{2},
4\right)$\\
    2 & $\left(\rep{1},\rep{1},\rep{1},\rep{1}\right)_{l}$ & $\left(
1,    8,  114,    5,    0,    5,   -5,   55,-\tfrac{199}{2},
-4\right)$\\
    2 & $\left(\rep{1},\rep{1},\rep{1},\rep{1}\right)_{l}$ &
$\left(-\tfrac{3}{2},   -8,   64,  -15,    0,  -15,   15,
115,-\tfrac{279}{2},    6\right)$\\
    2 & $\left(\rep{1},\rep{1},\rep{1},\rep{1}\right)_{l}$ & $\left(
3,   -5,  -49,   21,    0,   21,  -21,    7,-\tfrac{135}{2},
-12\right)$\\
    2 & $\left(\rep{1},\rep{1},\rep{1},\rep{1}\right)_{l}$ &
$\left(\tfrac{1}{4},-\tfrac{5}{2},-\tfrac{49}{2},  112,
0,\tfrac{19}{2},   94,
-278,\tfrac{245}{2},-\tfrac{119}{2}\right)$\\

    2 & $\left(\rep{1},\rep{1},\rep{1},\rep{1}\right)_{l}$ &
$\left(\tfrac{1}{4},\tfrac{5}{2},-\tfrac{307}{2}, -102,
0,\tfrac{1}{2}, -104,  108,\tfrac{233}{2},\tfrac{115}{2}\right)$\\
    2 & $\left(\rep{1},\rep{1},\rep{1},\rep{1}\right)_{l}$ &
$\left(\tfrac{9}{4},\tfrac{11}{2},\tfrac{179}{2},  -86,
0,\tfrac{33}{2}, -120,   60,\tfrac{297}{2},\tfrac{99}{2}\right)$\\
    2 & $\left(\rep{1},\rep{1},\rep{1},\rep{1}\right)_{l}$ & $\left(
1,-\tfrac{5}{2},-\tfrac{49}{2},  112,    0,\tfrac{19}{2}, -113,
81,\tfrac{269}{2},-\tfrac{125}{2}\right)$\\
    2 & $\left(\rep{1},\rep{2},\rep{1},\rep{1}\right)_{l}$ &
$\left(\tfrac{1}{4},-\tfrac{13}{2},\tfrac{15}{2},    3,    0, -202,
-5,  -85,\tfrac{239}{2},   -1\right)$\\
    2 & $\left(\rep{1},\rep{1},\rep{1},\rep{1}\right)_{l}$ &
$\left(-\tfrac{3}{4},\tfrac{13}{2},-\tfrac{15}{2},  200,    0, -210,
3,  -61,\tfrac{207}{2},    3\right)$\\
    2 & $\left(\rep{1},\rep{1},\rep{1},\rep{1}\right)_{l}$ &
$\left(-\tfrac{1}{4},\tfrac{13}{2},-\tfrac{15}{2}, -206,    0, -206,
-1,  -73,\tfrac{223}{2},    1\right)$\\
    2 & $\left(\rep{1},\rep{1},\rep{1},\rep{1}\right)_{l}$ &
$\left(\tfrac{3}{4},    4,  -32,  109,    0,\tfrac{423}{2},   99,
-123,-\tfrac{213}{2},\tfrac{111}{2}\right)$\\
    2 & $\left(\rep{1},\rep{1},\rep{1},\rep{2}\right)_{l}$ &
$\left(\tfrac{5}{4},    4,  -32,  109,    1,\tfrac{13}{2},   28,
84,\tfrac{265}{2},\tfrac{107}{2}\right)$\\
    2 & $\left(\rep{1},\rep{1},\rep{1},\rep{1}\right)_{l}$ &
$\left(\tfrac{5}{4},    4,  -32,  109,   -2,\tfrac{13}{2},   28,
84,\tfrac{265}{2},\tfrac{107}{2}\right)$\\
    2 & $\left(\rep{1},\rep{2},\rep{1},\rep{1}\right)_{l}$ &
$\left(\tfrac{5}{4},-\tfrac{5}{2},\tfrac{129}{2},  -96,   -1,
-96,\tfrac{123}{2},-\tfrac{51}{2},   17,\tfrac{107}{2}\right)$\\
    2 & $\left(\rep{1},\rep{1},\rep{1},\rep{1}\right)_{l}$ &
$\left(\tfrac{9}{4},-\tfrac{5}{2},-\tfrac{227}{2},  117,   -1,
-88,\tfrac{107}{2},-\tfrac{99}{2},   33,\tfrac{99}{2}\right)$\\
    2 & $\left(\rep{1},\rep{1},\rep{1},\rep{1}\right)_{l}$ &
$\left(-\tfrac{7}{4},\tfrac{5}{2},-\tfrac{129}{2}, -107,    1,
98,-\tfrac{127}{2},-\tfrac{241}{2},  206,-\tfrac{103}{2}\right)$\\
    2 & $\left(\rep{1},\rep{1},\rep{1},\rep{1}\right)_{l}$ &
$\left(\tfrac{1}{4},\tfrac{21}{2},\tfrac{99}{2},  101,   -1,
-104,\tfrac{139}{2},-\tfrac{3}{2},    1,\tfrac{115}{2}\right)$\\
    2 & $\left(\rep{1},\rep{1},\rep{1},\rep{1}\right)_{l}$ &
$\left(-\tfrac{1}{2},\tfrac{5}{2},-\tfrac{129}{2}, -107,   -1,
-107,\tfrac{145}{2},\tfrac{15}{2},   -5,-\tfrac{113}{2}\right)$\\
    2 & $\left(\rep{1},\rep{1},\rep{3},\rep{1}\right)_{l}$ &
$\left(\tfrac{5}{4},\tfrac{3}{2},\tfrac{65}{2},   13,
-1,-\tfrac{179}{2},-\tfrac{97}{2},-\tfrac{291}{2}, 97,   -5\right)$\\
    2 & $\left(\rep{1},\rep{1},\rep{1},\rep{1}\right)_{l}$ &
$\left(\tfrac{5}{4},\tfrac{3}{2},\tfrac{65}{2},   13,
-1,\tfrac{231}{2},\tfrac{321}{2},-\tfrac{437}{2},20,   -5\right)$\\
    2 & $\left(\rep{1},\rep{1},\rep{1},\rep{2}\right)_{l}$ &
$\left(\tfrac{3}{4},\tfrac{3}{2},\tfrac{65}{2},   13,
0,\tfrac{231}{2},-\tfrac{231}{2},-\tfrac{413}{2}, 12,   -3\right)$\\
    2 & $\left(\rep{1},\rep{1},\rep{1},\rep{1}\right)_{l}$ & $\left(
2,\tfrac{3}{2},\tfrac{65}{2},   13,
-1,\tfrac{231}{2},-\tfrac{93}{2},\tfrac{281}{2},   32,   -8\right)$\\
    2 & $\left(\rep{1},\rep{1},\rep{1},\rep{1}\right)_{l}$ & $\left(
2,\tfrac{3}{2},\tfrac{65}{2},   13,
1,-\tfrac{179}{2},\tfrac{41}{2},-\tfrac{157}{2}, -199,   -8\right)$\\
    2 & $\left(\crep{3},\rep{1},\rep{1},\rep{1}\right)_{l}$ &
$\left(-\tfrac{1}{4},   -4,  -57, -104,    1,
-104,-\tfrac{137}{2},\tfrac{9}{2},   -3,-\tfrac{37}{2}\right)$\\
    2 & $\left(\rep{1},\rep{1},\rep{1},\rep{1}\right)_{l}$ &
$\left(\tfrac{3}{4},    4,   57,  104,   -1,
104,\tfrac{137}{2},\tfrac{271}{2}, -216,-\tfrac{123}{2}\right)$\\
    2 & $\left(\rep{1},\rep{1},\rep{1},\rep{1}\right)_{l}$ &
$\left(\tfrac{13}{4},   -1,    8,  -78,    1,
-78,-\tfrac{189}{2},-\tfrac{147}{2},   49,\tfrac{91}{2}\right)$\\
    2 & $\left(\rep{1},\rep{1},\rep{1},\rep{1}\right)_{l}$ & $\left(
0,    4,   57,  104,    1, -101,-\tfrac{143}{2},-\tfrac{9}{2},
3,-\tfrac{117}{2}\right)$\\
\end{longtable}

\end{document}